\documentclass[superscriptaddress,aps,prl,showkeys,showpacs,twocolumn,10pt]{revtex4}
\usepackage[OT1,T1]{fontenc}
\usepackage{graphicx}
\usepackage{rotating}
\begin{document}

\title{On a Possible Explanation of the DLS-Puzzle}
\date{\today}

\newcommand*{\PITue}{Physikalisches Institut, Eberhard--Karls--Universit\"at
  T\"ubingen,  
%Auf der Morgenstelle~14, 
72076 T\"ubingen, Germany}

\author{M.~Bashkanov}   \affiliation{\PITue}
\author{H.~Clement}     \affiliation{\PITue}

\begin{abstract}
The enhancement in the dilepton spectrum observed in heavy-ion collisions for
invariant electron-positron masses in the range 0.15 GeV/c$^2 < M_{e^+e^-} <$ 0.6
GeV/c$^2$ has recently been traced back to a corresponding enhancement in
$pn$ collisions relative to $pp$ collisions. Whereas the dilepton spectra in the
latter are understood quantitatively, theoretical descriptions fail to describe
the much higher dilepton rate in $pn$ collisions, in particular regarding the
region  $M_{e^+e^-} >$ 0.3 GeV/c$^2$ at beam energies below 2 GeV. We show
that the missing strength can be 
attributed to the $\rho$-channel $\pi^+\pi^-$-production, which is dominated
by the $t$-channel $\Delta\Delta$ excitation and the recently found isoscalar
dibaryonic resonance structure at 2.37~GeV. 
\end{abstract}


\maketitle

\section{Introduction}

Dilepton spectroscopy has been established as a valuable tool to explore the
conditions of matter at high temperature and high density. Such extreme
conditions as found in stars or in the early universe can be probed by
relativistic heavy-ion collisions. In measurements of such collision processes
a significant excess of lepton pairs over the theoretically expected rate has
been observed in the mass region between the pion mass and the $\omega$ mass
%("DLS puzzle") \cite{dls} 
and interpreted as a possible sign of medium
modifications. However, at lower beam energies of (1 - 2) GeV per nucleon still
such an enhancement has been observed.

To address this problem the Dilepton Spectrometer (DLS) collaboration was the
first to investigate the underlying basic reactions by studying the dilepton
production in n$pp$ and $pn$ collisions. As a result they found that an
enhancement persists even for beam energies as low as 1 GeV ("DLS puzzle")
\cite{dls}. 
In recent measurements of dilepton pairs produced in C~+~C, $p + n$ and $p + p$
collisions at HADES the enhancement observed in heavy-ion collisions could be
traced back to such a one in $pn$ relative to $pp$ collisions processes
\cite{hades}. A
number of theoretical calculation have been successful in explaining the
dilepton spectrum originating from $pp$ collisions. They also succeed in
predicting a significantly higher dilepton rate for $pn$ collisions. However,
they all %largely 
under-predict the $pn$ induced dilepton production for
$M{e^+e^-} >$ 0.3 GeV/c$^2$ by up to an order of magnitude
\cite{Mosel1, Mosel2, Mart,brat} at beam energies below 2 GeV, 
      though Ref. \cite{Mosel1} can cure much of the disagreement by
      introduction of a pion electromagnetic form-factor. 

In these calculations the following lepton-pair production processes have been
taken into account:
\begin{itemize}
\item pion Dalitz decay $\pi^0 \to e^+e^-\gamma$,
\item $\eta$ Dalitz decay $\eta \to e^+e^-\gamma$,
           \item leptonic vector meson decay $v \to e^+e^-$,
\item virtual bremsstrahlung $NN \to NN e^+e^-$ and
\item baryon resonance decay $R \to N e^+e^-$, predominantly $\Delta \to N
  e^+e^-$.  
\end{itemize}

   The bremsstrahlung calculations of Ref. \cite{KK} overshoot the HADES data
   for $pp$ collisions. For the $np$ case they overshoot the data for
   $M_{e^+e^-}<$ 0.3 GeV and underpredict them above.

At sufficiently high incident energies both colliding nucleons may get
excited. So in addition to the configuration $NR$, where $R$ denotes a nucleon
in one of its excited states (resonance), we may have combinations of the form
RR'. The 
lowest-lying such configuration is $\Delta\Delta$. In the following we will
concentrate on the beam energy 1.25 GeV, where high-precision HADES data are
available. As we will argue below the only relevant RR' configuration there is 
$\Delta\Delta$.

At the energies of interest here, single-pion production in $NN$ collisions is
by far the largest inelastic channel. It is
dominated by $t$-channel meson exchange in combination with the excitation of
one of the nucleons into the $\Delta(1232)$ resonance - or to a lesser extent
to the Roper resonance $N^*(1440)$ with subsequent decay into the $\pi N$
system. 

In the description of the dilepton spectra two-pion production has not been
taken into account 
      in most of the previous works, 
since its cross section is smaller by a order of
magnitude. However, as we will show in the following, due to the relatively
large decay branching $\rho^0 \to e^+e^-$ the $\pi^+\pi^-$ production in the
$\rho$ channel contributes significantly to the electron-positron spectrum for
$M_{e^+e^-} \geq $ 0.3 GeV/c$^2$. 
 
   In Ref. \cite{Mosel2} two-pion production
   has been accounted for in some global manner. Here we proceed
   differently. Since the two-pion channels have been investigated
   experimetally meanwhile by exclusive and kinematically complete
   measurements, we know the dominating two-pion production mechanisms in
   dependence of the energy in detail. In particular, we may perform an isospin
   decomposition of experimental cross sections and underlying reaction
   mechanisms, in order to separate their contributions to $pp$ and $pn$
   induced dilepton production.

\section{Two-pion production}

In recent years the two-pion production in $pp$ and $pn$ collisions has been
measured by exclusive and kinematically complete experiments over the energy
region from threshold up to $T_{lab}$ = 1.4 GeV
\cite{JJ,WB,JP,TS,iso,FK,deldel,nnpipi,tt,AE,evd,mb,MB,isoabc,ts}.  

It has been shown that the
$pp$ induced two-pion production process is dominated by $t$-channel Roper and
$\Delta\Delta$ excitation
\cite{JJ,WB,JP,TS,iso,FK,deldel,nnpipi,tt,AE,evd,luis,xu}. 
In the latter both nucleons are mutually excited
to the $\Delta$ resonance by $t$-channel meson exchange in the collision
process. The Roper excitation process dominates at energies close to
threshold below 1 GeV, whereas the $\Delta\Delta$ process takes over above 1
GeV. Hence in the following we will focus on the latter two-pion
production process. And since the HADES experiment has been carried out at
$T_p$ = 1.25~GeV, we will concentrate on this energy.

In $pn$ induced two-pion production in addition the recently discovered dibaryon
resonance structure $d^*$ with $I(J^P) = 0(3^+)$, M = 2.37 GeV/c$^2$ and
$\Gamma$ = 70 MeV strongly contributes at energies around 1.2 GeV due to its
decay $d^* \to \Delta\Delta \to NN\pi\pi$ \cite{mb,MB,isoabc,ts}.

The total inclusive cross section for $pp$ induced $\pi^+\pi^-$ production at
$T_p$ = 1.25 GeV is about 700 $\mu$b and for $np$ induced $\pi^+\pi^-$
production it is about 1300 $\mu$b. The latter contains not only the
$np\pi^+\pi^-$ channel, but also the double-pionic fusion channel $d\pi^+\pi^-$.

The only sizeable way two-pion production may feed the electron-pair
production is via 
 $\pi^+\pi^- \to \rho^0 \to e^+e^-$ with the isovector $\pi^+\pi^-$
pair being in relative $p$-wave ($\rho$ channel).

In order to filter out the $\rho$-channel $\pi^+\pi^-$ production from the
known two-pion production cross sections, we make use of the isospin
decomposition of these cross sections in terms of matrix elements $M_{I_{NN}^f
  I_{\pi\pi}I_{NN}^i}$, where $I_{\pi\pi}$ stands for the isospin of the pion pair
and $I_{NN}^i$ and $I_{NN}^f$ for the isospin of the nucleon pair in initial and
final states, respectively \cite{dakhno,bys,iso}.

    For a specific process these matrix elements depend on the isospin
    coupling coefficients. For the $\Delta\Delta$ process the matrix elements
    are proportional to the respective 9j-symbol for isospin recoupling: 
\begin{eqnarray}
M_{I_{NN}^fI_{\pi\pi}I_{NN}^i}^{\Delta\Delta} \sim \hat{I_{\Delta_1}}\hat{I_{\Delta_2}}
%((I_{N_1}I_{\pi_1})I_{\Delta_1},(I_{N_2}I_{\pi_2})I_{\Delta_2},I_{\Delta\Delta}|(I_{N_1}I_{N_2})I_{NN},(I_{\pi_1}I_{\pi_2})I_{\pi\pi},I_{\Delta\Delta}\nonumber\\
\hat{I_{NN}}\hat{I_{\pi\pi}}
\left\{
\begin{array}{ccc}
I_{N_1}&I_{\pi_1}&I_{\Delta_1}\\
I_{N_2}&I_{\pi_2}&I_{\Delta_2}\\
I_{NN}&I_{\pi\pi}&I_{\Delta\Delta}\\
\end{array}
\right\},
\end{eqnarray}
      where $N_i$ and $\pi_i$ couple to $\Delta_i$ for $i = 1,2$ and
      $\hat{I_\alpha} = \sqrt{2I_\alpha + 1}$ .

In $pp$-initiated two-pion production only $M_{111}$ gives rise to
$\rho^0$-channel 
production. However, because  $M_{111}^{\Delta\Delta} \equiv$~0 for the
$\Delta\Delta$ process 
      --- since the corresponding 9j-symbol in eq. (1) is zero,
there is no contribution to $\rho^0$-channel production. Hence the PLUTO
\cite{pluto} generated cocktail for the description of the $pp$ dilepton
production as given in Ref. \cite{hades} stays unchanged.

%For the  Roper excitation process we have $M_{111}^{N^* \to \Delta\pi} = \frac
%1 2 M_{101}^{N^* \to \Delta\pi}$ \cite{iso}. Accounting for the branching
%ratio of $\Gamma(N^* \to 
%\Delta\pi) / \Gamma(N^* \to N\sigma) \approx$ 1 \cite{PDG,BoGa,Roper} we end
%up with $\sigma_{111}^{N^*} \approx \frac 1 {25} \sigma_{101}^{N^*}$. The
%latter is given in Ref.~\cite{iso} to be in the order of a few $\mu$b, which
%means that the $\rho^0$-channel $\pi^+\pi^-$ production cross section
%originating from the Roper excitation process is much below 1 $\mu$b. And since 
%$M_{111}^{\Delta\Delta}$~=~0 the $\rho^0$-channel production is only a tiny
%fraction of $pp$-initiated two-pion production.

The situation changes dramatically in case of $pn$-initiated $\rho^0$-channel
production, since here we indeed do have large contributions from the
$\Delta\Delta$-process.
According to Refs. \cite{dakhno,bys} we have for the $pn$ initiated
$\pi^+\pi^-$ production:
\begin{eqnarray}
\sigma(pn \to pn\pi^+\pi^-) = &&\\
=  &&\frac 1 {60}|\sqrt 5 M_{101} -M_{121}|^2 +\nonumber\\
+  &&\frac 1 8 |M_{011}|^2 + \frac 1 {24} |M_{110}|^2 +\nonumber\\
+  &&\frac 1 {12} |M_{000}|^2 \nonumber
\end{eqnarray}
and since $I_d$ = 0
\begin{eqnarray}
\sigma(pn \to d\pi^+\pi^-) = &&\\
=  &&\frac 1 8 |M_{011}|^2 +  \frac 1 {12} |M_{000}|^2. \nonumber
\end{eqnarray}

For dilepton production via $\rho^0$ production only matrix elements with
$I_{\pi\pi}$ = 1 contribute. Selecting in addition the $\Delta\Delta$ process
we end up with:

\begin{eqnarray}
\sigma(pn \to \Delta\Delta \to pn[\pi^+\pi^-]_{I=1})&& = \\
 =  &&\frac 1 8
|M_{011}^{\Delta\Delta}|^2 + \frac 1 {24} |M_{110}^{\Delta\Delta}|^2 \nonumber
\end{eqnarray}
and
\begin{eqnarray}
\sigma(pn \to \Delta\Delta \to d[\pi^+\pi^-]_{I=1}) = \frac 1 8
|M_{011}^{\Delta\Delta}|^2.  
\end{eqnarray}

With the relations
\begin{equation}
M_{011}^{\Delta\Delta} = \sqrt{\frac {15}9} M_{110}^{\Delta\Delta} = \sqrt{\frac {15} 2} M_{121}^{\Delta\Delta}
\end{equation}
obtained by angular momentum recoupling according to eq. (1)
this leads to 
\begin{eqnarray}
\sigma(pn \to \Delta\Delta \to pn[\pi^+\pi^-]_{I=1}) =  \frac {27} {16}
|M_{121}^{\Delta\Delta}|^2 =  \\
 = \frac {45} 4 \sigma(pp \to \Delta\Delta \to nn\pi^+\pi^+), \nonumber
\end{eqnarray}
since \cite{dakhno,bys}
\begin{eqnarray}
\sigma(pp \to nn\pi^+\pi^+) =  \frac {3} {20}|M_{121}|^2.
\end{eqnarray}
The analysis of the $pp \to nn\pi^+\pi^+$ reaction gives about 15 $\mu$b
\cite{nnpipi} for this cross section at $T_p$ = 1.25 GeV, which results in  
\begin{equation}
\sigma(pn \to \Delta\Delta \to pn[\pi^+\pi^-]_{I=1}) \approx 170 \mu b.
\end{equation}

This number roughly corresponds to one fourth of the full $\Delta\Delta$
production in the $pn \to pn\pi^+\pi^-$ reaction \cite{luis}.

In case the final pn pair fuses to a deuteron we also obtain $\rho^0$-channel
production, which is related to the measured $\pi^+\pi^0$ ($\rho^+$ channel)
production in $pp$ collisions by the isospin relation \cite{isoabc}:
 
\begin{eqnarray}
\sigma(pn \to  d[\pi^+\pi^-]_{I=1}) =  \frac {1} 2
\sigma(pp \to  d\pi^+\pi^0) \approx 100 \mu b.
\end{eqnarray}

In addition, the $\rho^0$-channel production in $pn$ initiated two-pion
production is fed by excitation and decay of the $d^*$-resonance. Since its
decay proceeds again via the $\Delta\Delta$ system in the intermediate step, we
can use the isospin relation for the $\Delta\Delta$ system -- utilizing again
eq. (1) --
\begin{equation}
M_{110}^{\Delta\Delta} = -\sqrt{\frac 1 2} M_{000}^{\Delta\Delta},
\end{equation}
in order to connect the $pn[\pi^+\pi^-]_{I=1}$ decay channel with the
$pn\pi^0\pi^0$ channel. According to the predictions in Refs. \cite{CW,EO} the
resonance effect in the latter channel should be 85 $\%$ of that in
$d\pi^0\pi^0$ channel \cite{isoabc}, which isospin-decomposed reads as
\cite{dakhno,bys} 
\begin{eqnarray}
\sigma(pn \to pn\pi^0\pi^0) =  &&\frac 1 {30}|\frac {\sqrt 5} 2 M_{101} + M_{121}|^2 +\\
+  &&\frac 1 {24} |M_{000}|^2 \nonumber
\end{eqnarray}
and for $I_{NN}^i = I_{d^*}$ = 0:
\begin{eqnarray}
\sigma(pn \to d^* \to d\pi^0\pi^0) = \frac 1 {24} |M_{000}|^2.
\end{eqnarray}

At the resonance maximum at $\sqrt s$ =
2.37 GeV this cross section is about 240 $\mu$b, however, at $\sqrt s$ = 2.42
GeV  ( $T_p$ = 1.25~GeV) it is already as low as  90~$\mu$b. Together with
eqs. (2), (11) and (13) and the condition $I_{NN}^i = I_{d^*}$ = 0 this results
in:
 
\begin{eqnarray}
\sigma(pn \to d^* \to pn[\pi^+\pi^-]_{I=1})&& = \\ 
 = && \frac 1 {24} |M_{110}^{\Delta\Delta}(d^*)|^2 = \nonumber \\ 
 = && \frac 1 {48} |M_{000}^{\Delta\Delta}(d^*)|^2  =  \nonumber \\ 
 = && \frac 1 2 \sigma(pn \to d^* \to pn\pi^0\pi^0) \nonumber \\
 \approx && 40 \mu b. \nonumber
\end{eqnarray}

A cross check of this number is provided by a recent measurement \cite{ts} of
the  $pp\pi^0\pi^-$ channel, since again by isospin relations we have
\cite{ts,dakhno,bys} 

\begin{eqnarray}
\sigma(pn \to d^* \to pn[\pi^+\pi^-]_{I=1}) =  \nonumber \\
  = \sigma(pn \to d^* \to pp\pi^0\pi^-).
\end{eqnarray} 

Though according to Gal and Garcilazo \cite{GG} the $d^*$ decay into isovector
nucleon and pion pairs should be
dynamically suppressed, the measurement of the $pn
\to d^* \to pp\pi^0\pi^-$ reaction and its analysis \cite{ts} is compatible
with a resonance cross section as expected by the isospin relations. However,
since in the $pp\pi^0\pi^-$ channel the resonance structure sits upon a large
background of conventional processes, it cannot be excluded that the resonance
contribution actually might be somewhat smaller.

In total we have about 310 $\mu$b of $\rho^0$-channel $\pi^+\pi^-$-production
in $pn$-initiated reactions at $T_p$ = 1.25 GeV --- compared to none in
$pp$-initiated reactions. We estimate this number to be correct at least
within 20 $\%$.

\section{$\rho^0$-channel $e^+e^-$ production}

To calculate the  $e^+e^-$ production we assume that the two pions produced in
the $\Delta\Delta$ process undergo final state interaction by forming a
$\rho^0$, which subsequently decays into a  $e^+e^-$ pair:
\begin{eqnarray}
pn \to \Delta\Delta \to pn[\pi^+\pi^-]_{I=L=1} \to pn\rho^0 \to pne^+e^-,
\end{eqnarray} 
see graphs in Fig.~1.
The intermediate $\Delta\Delta$ system is formed either by $t$-channel
meson exchange or by decay of the $d^*$ resonance with cross sections as
evaluated above. 

For the transition from the
$[\pi^+\pi^-]_{I=L=1}$ system into the $[e^+e^-]_{L=0}$ system by rescattering
(final state interaction in the $ \rho$-channel) we use a 
Breit-Wigner ansatz 
%for the $\rho^0$ propagator: as given by Li, Ko and Brown 
\cite{li,koch}:
\begin{eqnarray}
%|\hat{\rho^0}|^2 \sim %
|{\mathcal M}(\pi^+\pi^- \to \rho^0 \to e^+e^-)|^2 = %\frac {12\pi} { q^2} 
\frac {m_{\rho}^2 \Gamma_{\pi^+\pi^-} \Gamma_{e^+e^-}}
{(s - m_{\rho}^2)^2 + m_{\rho}^2\Gamma_{\rho}^2}. 
\end{eqnarray} 

For the p-wave decay into the $\pi^+\pi^-$ channel we have $\Gamma_{\pi^+\pi^-}
\sim q^3$ and for the s-wave decay into the $e^+e^-$ channel we have
$\Gamma_{e^+e^-} \sim k$, where $q$ and
$k$ are the momenta in $\pi^+\pi^-$ and $e^+e^-$ subsystems, respectively. In
a more detailed consideration \cite{koch} the partial widths depend also on the
invariant masses $M_{\pi^+\pi^-}$ and $M_{e^+e^-}$ yielding
$\Gamma_{\pi^+\pi^-} = a q^3/M_{\pi^+\pi^-}$ and $\Gamma_{e^+e^-} = b k /
M_{e^+e^-}^3$. The  
constants a and b in the partial widths are fixed by adjusting them to the
known branching ratios and widths at the $\rho$ mass pole \cite{PDG}. Hence the
Monte Carlo (MC) simulation of process~(16) is straightforward and free of
parameters. 

\begin{figure}
\centering
\includegraphics[width=0.99\columnwidth]{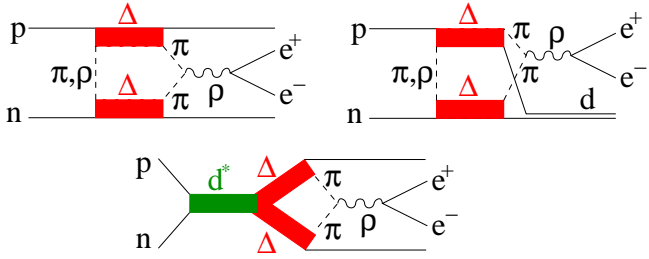}
\caption{\small (Color Online) Graphs for the $e^+e^-$ production via $\rho^0$
  channel 
  $\pi^+\pi^-$ production in $pn$ collisions. Top: production via $t$-channel
  $\Delta\Delta$ excitation leading to $pn$ (left) and deuteron (right)
  final states. Bottom: production via $s$-channel $d^*$ formation and its
  subsequent decay into the $\Delta\Delta$ system.
}
\label{fig1}
\end{figure}

\section{Results}

\begin{figure}
\centering
\includegraphics[width=0.99\columnwidth]{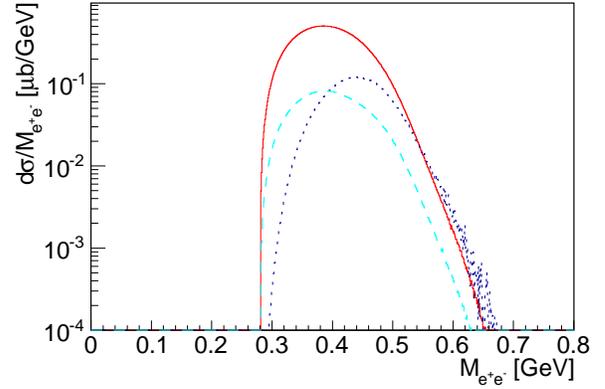}
\includegraphics[width=0.99\columnwidth]{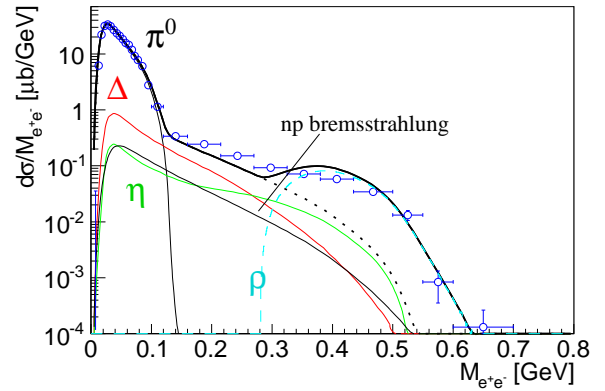}
\caption{\small (Color online) Distribution of the invariant mass $M_{e^+e^-}$
  produced in 
  $pn$ collisions at $T_p$ = 1.25 GeV. Top: $e^+e^-$ production from $\rho^0$
  decay resulting from the $\Delta\Delta$ excitation via on-shell $\pi^+\pi^-$
  production according to
  process~(16). The drawn curves denote the $[\pi^+\pi^-]_{I=J=1}$ spectrum
  scaled by the $e^+e^-$ branching ratio at the pole of $\rho^0$ (blue
  dotted), the resulting $e^+e^-$ spectrum  using the proper momentum dependent
  branching ratio (red solid) and the resulting $e^+e^-$ spectrum within the
  HADES acceptance (cyan dashed). Bottom:  Full $e^+e^-$ production. The open
  circles give the HADES result \cite{hades}. Thin solid lines denote
  calculations for $e^+e^-$ production originating from $\pi^0$ production and
  bremsstrahlung (black), single $\Delta$ (red) and $\eta$ (green)
  production with subsequent Dalitz decay. The dotted curve denotes
  the sum of these processes. The dashed (cyan) curve gives the contribution
  from the $\rho^0$-channel $\pi^+\pi^-$ production and the thick solid line
  the sum of all these processes.
}
\label{fig1}
\end{figure}

The numerical results of this MC simulation are displayed in Fig.~2. At the
top panel we
show first the $\rho^0$-channel $\pi^+\pi^-$ spectrum obtained from the
processes discussed in eqs. (1) - (15) and scaled by the $e^+e^-$ branching
ratio at the pole of $\rho^0$ (dotted line). This gives only a crude
estimate. A proper treatment involves the momentum-dependent transition
amplitude in eq. (17) resulting in the solid curve. The enhanced
yield of the $e^+e^-$ spectrum relative to the scaled $\pi^+\pi^-$ spectrum at
low masses is due to the fact that -- in addition to the inverse power
dependence on the invariant mass -- the pion pair is in relative $p$-wave and
therefore suppressed near threshold, whereas the  $e^+e^-$  pair is in relative
$s$-wave and hence not suppressed. The resulting integral cross section for
the process $pn \to e^+e^-X$ is 72 nb, which is about a factor of four larger
than that from the crude estimate. 

Since the HADES detector has limited acceptance, this has to be taken into
account for 
comparison with the HADES data. The dashed curve exhibits the final
$e^+e^-$ production resulting from $\rho^0$-channel $\pi^+\pi^-$ production in
$pn$ collisions within the HADES acceptance.

All other conventional processes due to $\pi^0, \eta$ and $\Delta$ Dalitz
decays and bremsstrahlung -- mentioned in the introduction -- were simulated
using the PLUTO generator \cite{pluto} and filtered with HADES
efficiency-acceptance filters. 
    \footnote{
    Note that in PLUTO the bremsstrahlung
    contributions are taken according to the presciption of Ref. \cite{Mosel1},
    which leads to a good description of the HADES data for the $pp$ case
    in contrast to that of Ref. \cite{KK}.}
%On Fig.1 top one can see the effect of efficiency-acceptance red solid
%curve compare to cyan dashed. Lower yeld in $\pi^+\pi^-$ compare to
%$e^+e^-$ comes from $P-wave$ suppression at threshold in pionic channel.
They are shown in Fig.~2, bottom in comparison with the HADES data for $pn$
initiated $e^+e^-$ production at $T_p$ = 1.25 GeV. The sum of these
processes resulting from Dalitz decays is denoted by the dotted curve. It
provides a quantitative description of the data in 
the region of the $\pi^0$ peak, {\it i.e.} for $M_{e^+e^-} <$ 0.15 GeV. Above,
the sum curve under-predicts the data increasingly with increasing $M_{e^+e^-}$
values. However, if we add the $e^+e^-$ production resulting from
$\rho^0$-channel $\pi^+\pi^-$ production (dashed curve both in top and
bottom parts of Fig.~2) we obtain a nearly perfect description of the HADES
data. 

There appears still a slightly underestimated region in the range  0.15
GeV $< M_{e^+e^-} <$ 0.3 GeV. It possibly might be related to direct $d^*$
decay $pn \to d^* \to de^+e^-$ or $pn \to d^* \to [pn]_{I=0}e^+e^-$ as
suggested in Ref. \cite{Mart}.  
%$pn \to d^* \to d[e^+e^-]_{I=0}$ or $pn \to d^* \to [pn]_{I=0}(e^+e^-)_{I=0}$. 
However, since we know neither shape nor strength of such a $d^*$ form-factor
in this channel, we cannot estimate such a contribution reliably. In addition,
also the PLUTO generated processes have theoretical uncertainties, which are
in the order of the deviation in question. 
   Since we base here the dilepton production due to $\rho^0$ channel
   $\pi^+\pi^-$ production on experimental results for the relevant two-pion
   production channels, we consider here only the on-shell situation. However,
   because the
   two-lepton threshold is much lower than the two-pion threshold also dilepton
   production via virtual $\rho^0$ formation in the intermediate state will
   contribute. Taking this into account removes the cut in the $e^+e^-$ spectrum
   at the $\pi^+\pi^-$ threshold and replaces it by a smooth continuation as
   depicted, {\it e.g.}, in Fig.~3 of Ref.~\cite{Mosel2}. Hence accounting for
   this will fill up the gap below 0.3 GeV -- possibly overshoot it even
   somewhat. We refrain here from doing such a calculation, since in contrast
   to the on-shell consideration pursued here the off-shell contribution is
   model-dependent.

Finally we shortly comment on the dependence of the $e^+e^-$ spectrum on the
beam energy. The DLS collaboration has measured the $e^+e^-$ 
production in $pp$ and $pd$ collisions at several beam energies between 1.04
and 4.88 GeV. The ratio R of integrated yields for $M_{e^+e^-} >$ 0.15 GeV/c$^2$
exhibits a peak-like structure with a substantial rise from R $\approx$ 6 to R
$\approx$ 9 between $T_p$ = (1.0 -
1.27) GeV, falling off thereafter by a factor of roughly three until $T_p$~=~2
GeV. % -- with supposedly less variation above this energy. 
At 2.1 GeV the ratio
is somewhat above 2 and at 4.9 GeV a little bit below 2. %--- see Fig.~2.

Assuming the pd collisions to proceed mainly as quasifree proton-nucleon
collisions, we expect a ratio of R = 2, if $pp$ and $pn$ collisions contribute
equally much. In the quasifree picture the peak region
corresponds to 2.3 GeV $< \sqrt s <$ 2.7 GeV with the maximum around $\sqrt s
\approx$ 2.4 GeV, {\it i.e.} just in the region, where both the $d^*$ resonance
formation and the $pn \to \Delta\Delta \to d [\pi^+\pi^-]_{I=1}$   process
peak. Whereas the first one with a width of 70 MeV fades away below $\sqrt s$
= 2.3 GeV and above $\sqrt s$ = 2.5 GeV, the latter one with a width of about
250 MeV \cite{FK,isoabc} declines much slower fading away above $\sqrt
s$ = 2.8 GeV, which corresponds to beam energies beyond 2~GeV.    

For beam energies beyond 1.5 GeV ($\sqrt s$ = 2.5 GeV) we face substantial
contributions from the $\rho^0$ decay of higher-lying $N^*$ and $\Delta$
resonances, which get excited during the collision process. These sources
contribute to the dilepton spectra both from $pn$ and $pp$ collisions as
demonstrated by Refs. \cite{Mosel2,brat}, who succeed in a
quantitative description of the data for beam energies of 2 GeV and beyond.
%similar strength (?). As in Ref. \cite{hades2GeV} we simulate them with the
%PLUTO  event generator.

%For the energy region $T_p$ = 1 - 2 GeV the cross sections for the processes $pn
%\to d^* \to pn\rho$ and $pn \to \Delta\Delta \to pn\rho$ are well known due to
%the isospin relations (5) and (6). For the estimation of the $pp \to
%\Delta\Delta \to nn\pi^+\pi^+$ cross section in eq. (3) beyond 1.5 GeV we use
%the calculation of Ref. \cite{xu}. 

%\begin{figure}
%\centering
%\includegraphics[width=0.99\columnwidth]{ratio.eps}
%\caption{\small (Color online) 
%  Ratio of $e^+e^-$ yields for $M_{e^+e^-} >$ 0.15 GeV/c$^2$ in
%  $pd$ and $pp$ collisions in dependence of the beam energy. The solid
%  symbols denote the results from the DLS collaboration; the error
%  bars are statistical only, whereas the brackets above and below the data
%  points include the effects of normalization uncertainties \cite{dls}. The
%  dotted line shows the result from the PLUTO event generator for the
%  traditional processes. The dashed (cyan) curve gives the contribution
%  from the $\rho^0$-channel $\pi^+\pi^-$ production and the thick solid line
%  the sum of all these processes.
%}
%\label{fig1}
%\end{figure}

\section{Conclusions}

It has been shown that the $e^+e^-$ production resulting from $\rho^0$-channel
$\pi^+\pi^-$ production gives significant contributions to the dilepton
spectrum for $M_{e^+e^-} > 2m_{\pi}$, which account very well for the missing
strength in previous interpretations offering thus a solution of the
long-standing DLS puzzle.

\section{Acknowledgments}

We  are grateful to Tetyana Galatyuk and Malgorzata Gumberidze for their help
with HADES data and filtering software. We also want to thank Piotr Salabura
for stimulating and fruitful discussions, in particular for drawing our
attention to this issue. We also acknowledge valuable discussions with Avraham
Gal,Thomas Gutsche, Christoph Hanhart, Janus Weil and Colin Wilkin.
This work has been supported by the Forschungszentrum J\"ulich (COSY-FFE).

\end{document}